\documentclass[aps,floatfix,twocolumn,preprintnumbers,amsmath]{revtex4}
\usepackage{amssymb}
\usepackage[T1]{fontenc}
\usepackage[latin1]{inputenc}
\usepackage{graphicx}
\usepackage{dcolumn}
\usepackage{bm}

\begin{document}

\title{Quark deconfinement and neutrino trapping in compact stars}
                                                                                                                                                                                                                 
\author{I. Vida\~na$^1$, I. Bombaci$^2$ and I. Parenti$^3$}
                                                                                                                                                                                                                 
\address{$^1$Gesellschaft f\"{u}r Schwerionenforschung (GSI). Planckstrasse 1,
D-64291 Darmstadt, Germany.}
                                                                                                                                                                                                                 
\address{$^2$Dipartimento di Fisica ``E. Fermi'', Universit\`a di Pisa and
INFN sezione di Pisa. Largo Bruno Pontecorvo 3, I-56127 Pisa, Italy.}
                                                                                                                                                                                                                 
\address{$^3$Dipartimento di Fisica, Universit\`a di Ferrara and INFN sezione di Ferrara. Via Paradiso 12,
I-44100 Ferrara, Italy.}

\date{\today}

\begin{abstract}

We study the role played by neutrino trapping on the hadron star (HS) to quark star (QS) conversion mechanism proposed recently by
Berezhiani and collaborators. We find that the nucleation of quark matter drops inside hadron matter, and therefore the conversion of a HS
into a QS, is strongly inhibit by the presence of neutrinos.

\end{abstract}

\maketitle


\section{Introduction}
                                                                                                                                                                                                                 
Nowadays it is still an open question which is the true nature of neutron stars. In a traditional picture the
core of a neutron star is modeled as an uniform fluid of neutron-rich matter in equilibrium with respect to the weak
interaction. Nevertheless, due to the large stellar central density new degrees of freedom such as hyperons, meson
condensates or even a deconfined phase of quark matter are expected to appear in the inner core of the star
\cite{bo03}. The latter possibility is a consequence of the QCD, that predicts a phase transition from hadron matter to a
deconfined quark phase at few times nuclear saturation density, and was realized by several researchers soon after
the introduction of quarks as the fundamentals building blocks of hadrons \cite{iva69}.
                                                                                                                                                                                                                 
Berezhiani {\it et al.} \cite{be03} have shown recently that when finite effects at the interface between the
quark and the hadron phase are taken into account, pure HS ({\it i.e.,} without a phase of deconfined
quarks), above a threshold value of the central pressure, are metastable to ``decay'' into a more compact star in
which deconfined quark matter is present (QS). The mean-life time of the metastable star configurations is
related to the time needed to form a drop of quark matter in the stellar center, and depends dramatically on
the stellar central pressure. In this work, we study the role played by neutrino trapping on this conversion mechanism.
The quantum nucleation of a quark matter drop inside hadron matter is briefly revised
in Sec.\ \ref{sec:sec2}. Our main results are presented in Sec.\ \ref{sec:sec3}, whereas the main conclusions are given
in Sec.\ \ref{sec:sec4}.


\section{Quantum nucleation of quark matter in hadron stars}
\label{sec:sec2}
Let us consider a pure HS whose central density (pressure) is increasing due to spin-down or due to
mass accretion (from a companion star or from the interstellar medium). As the central density approaches the
quark deconfinement critical density, a drop of quark matter (QM) can be formed in the central region of the star.
The process of formation of the drop is regulated by its quantum fluctuations in the potential well created
from the difference between the energy densities of the hadron and quark phases. Keeping only
the volume and the surface terms, the potential well takes the simple form
\begin{equation}
U(R)=\frac{4}{3}\pi n_Q(\mu_Q-\mu_H)R^3 + 4\pi \sigma R^2
\label{eq:potential}
\end{equation}
where $n_Q$ is the quark baryon number density, $\mu_Q$ and $\mu_H$ are the quark and hadron chemical potentials
at a fixed pressure $P$, and $\sigma$ is the surface tension for the surface separating the hadron from the quark
phase.

The time needed to form the drop (nucleation time) can be straightforwardly evaluated within a semiclassical
approach \cite{iida97} . First one computes, in the Wentzel--Kramers--Brillouin (WKB) approximation, the ground state
energy $E_0$ and the oscillation frequency $\nu_0$ of the drop in the potential well $U(R)$. Then, the probability of
tunneling is given by
\begin{equation}
p_0=exp\left[-\frac{A(E_0)}{\hbar}\right]
\label{eq:prob}
\end{equation}
where $A$ is the action under the potential barrier which in a relativistic framework reads
\begin{equation}
A(E)=\frac{2}{c}\int_{R_-}^{R_+}\sqrt{[2M(R)c^2+E-U(R)][U(R)-E]} \ .
\label{eq:action}
\end{equation}
being $R_\pm$ the classical turning points and
\begin{equation}
M(R)=4\pi \rho_H\left(1-\frac{n_Q}{n_H}\right)^2R^3
\label{eq:mass}
\end{equation}
the droplet effective mass, with $\rho_H$ and $n_H$ the hadron energy density and the hadron baryon number
density, respectively. The nucleation time is then equal to
\begin{equation}
\tau=(\nu_0 p_0 N_c)^{-1} \ ,
\label{eq:time}
\end{equation}
where $N_c$ is the number of virtual centers of droplet formation in the star. A simple estimation gives $N_c \sim
10^{48}$ \cite{iida97}.
\begin{figure}[h]
\vspace{0.325cm}
\centerline{
     \includegraphics[scale=0.30]{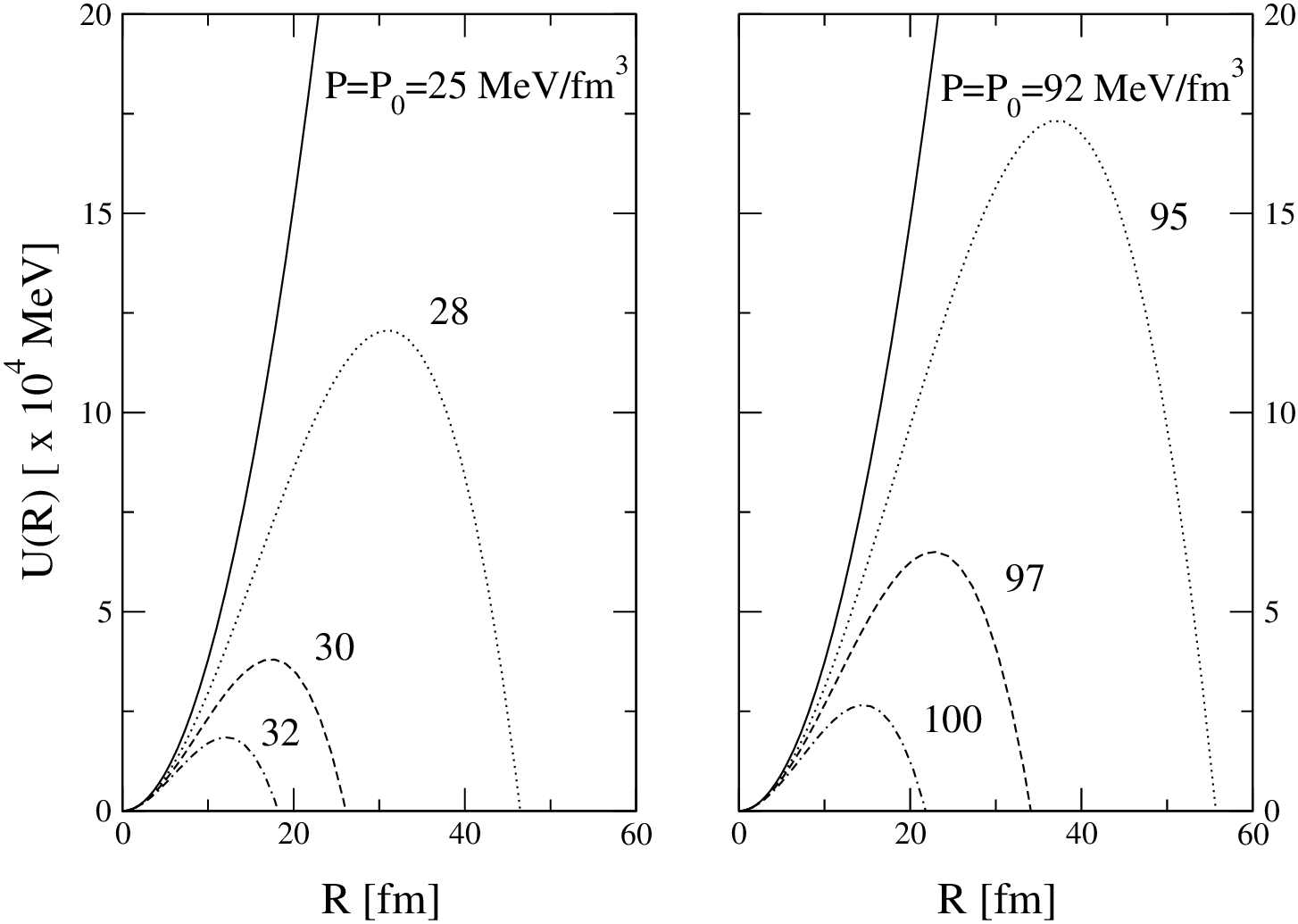}}
      \caption{Potential energy U(R) of the QM drop as a function of the radius of the drop.
      Left (right) panel corresponds to the neutrino free (trapped) case.}
        \label{fig:figure1}
\end{figure}
To describe both the hadron and the quark phases we have adopted rather common models. The hadron phase is
described within the relativistic mean field approximation with standard parameters fitted to reproduce nuclear matter
saturation properties \cite{gm91}, whereas for the quark phase we have adopted a phenomenological Equation of State
(EoS) based on the MIT bag model for hadrons \cite{fa84} with three parameters: the mass of the strange quark $m_s$,
the so-called pressure of the vacuum $B$ (bag constant) and the QCD structure constant $\alpha_c$. In the present
work, we take $m_u=m_d=0, m_s=150$ MeV and $\alpha_c=0$ (see Ref.\ \cite{bo04} for details).


\section{Results}
\label{sec:sec3}
                                                                                                                                                                                                                 
\begin{figure}[h]
\vspace{0.325cm}
\centerline{
     \includegraphics[scale=0.30]{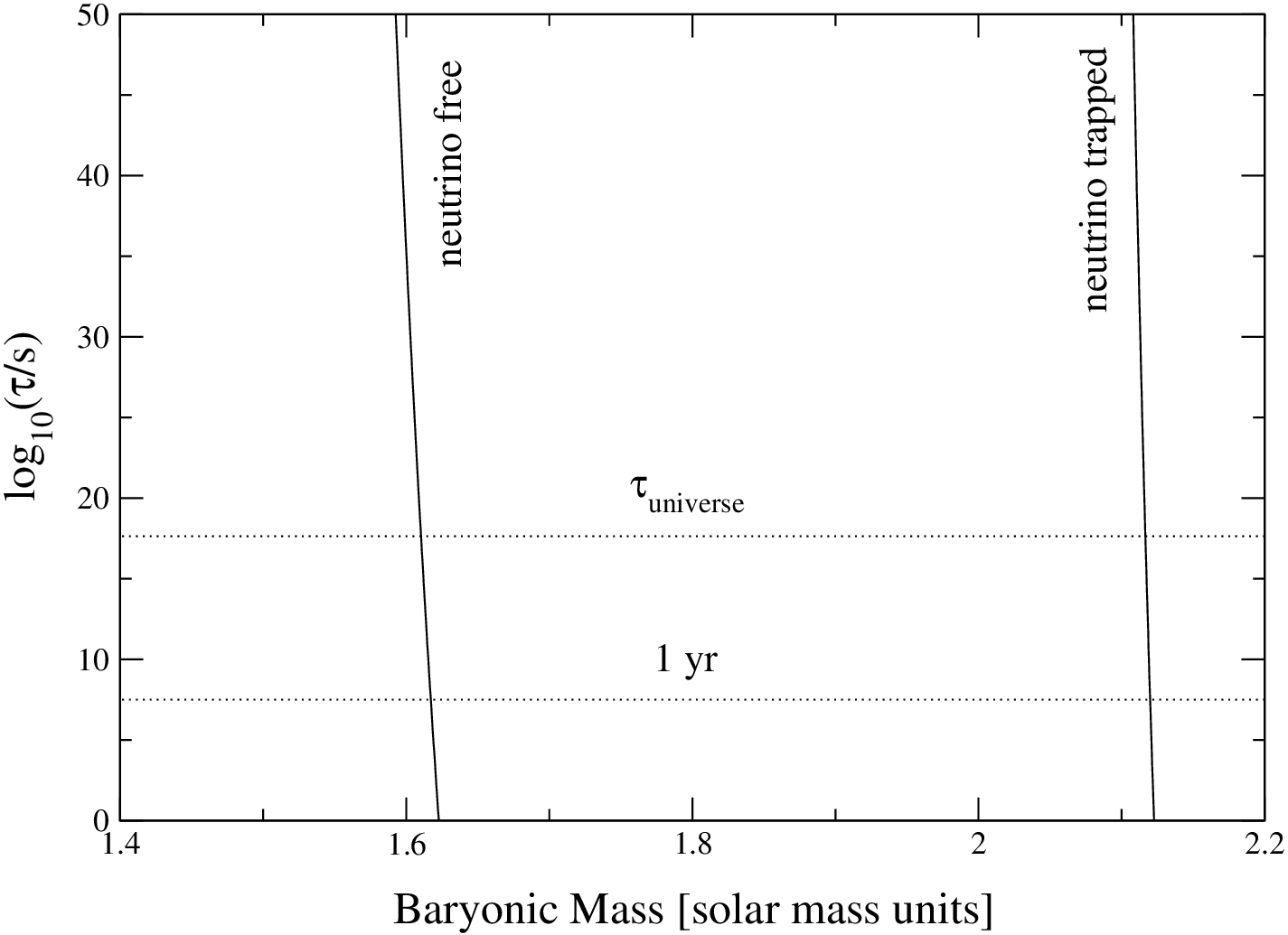}}
      \caption{Nucleation time as a function of the baryonic mass.}
        \label{fig:figure2}
\end{figure}

\begin{figure}[h]
\vspace{0.325cm}
\centerline{
     \includegraphics[scale=0.30]{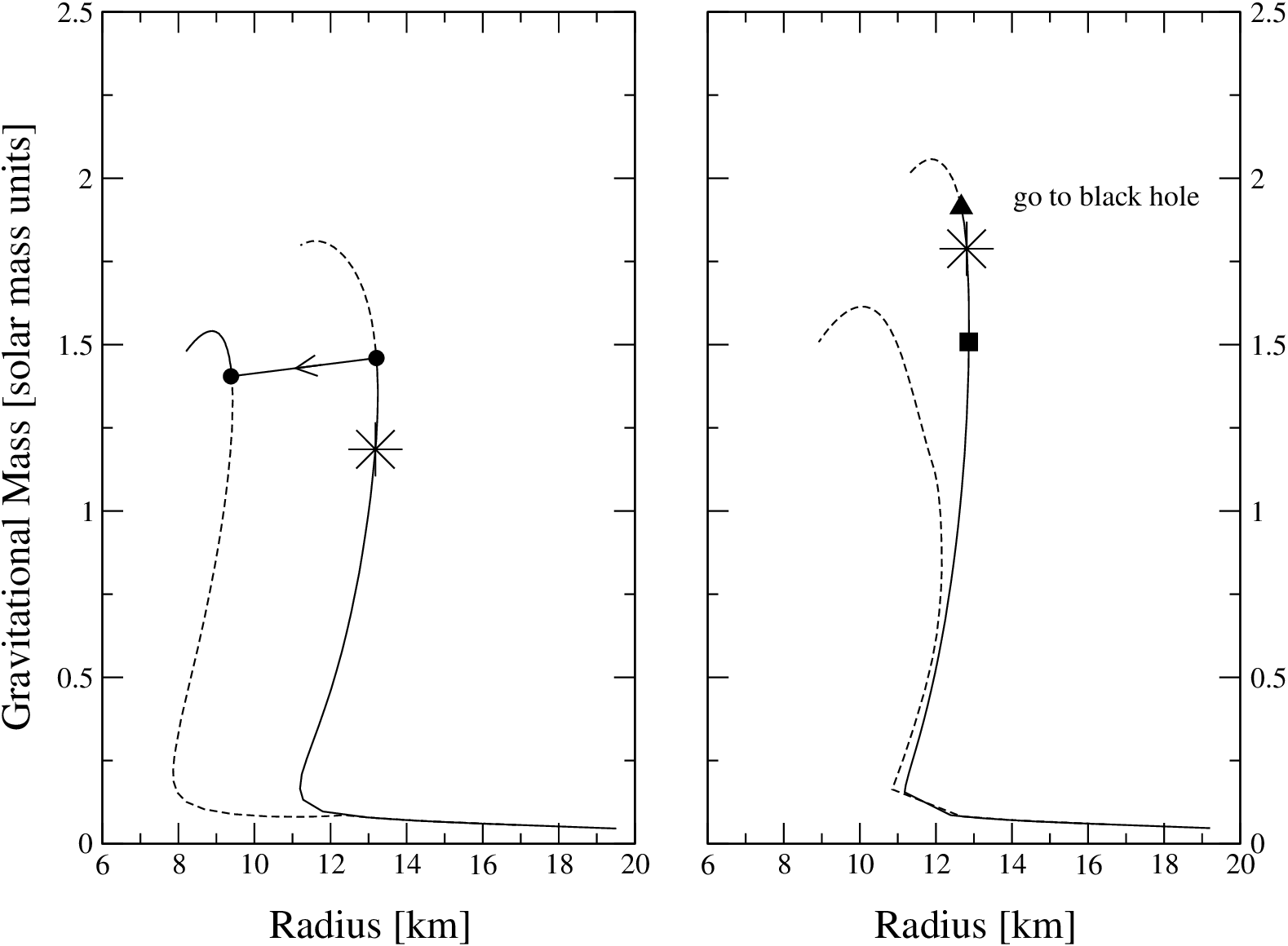}}
      \caption{Mass-radius relation for the initial HS and final QS configurations.
      Left (right) panel corresponds to the neutrino free (trapped) case.}
        \label{fig:figure3}
\end{figure}

To begin with we show in Fig.\ \ref{fig:figure1} the potential energy $U(R)$ for the formation of a QM
drop for different values of the stellar central pressure above the so-called static transition point $P_0$ (the
pressure at which the hadron and quark chemical potentials are equal). The surface tension $\sigma$ is taken equal to
$30$ MeV/fm$^2$ and $B=85.29$ MeV/fm$^3$. Right (left) panel corresponds to the case in which the presence of
neutrinos has (not) been taken into account in the EoS of the hadron and quark phases. As expected the potential barrier is
lowered as the central pressure increases. Note that the static transition point $P_0$ is larger when neutrinos are
present, and note also that in this case, higher values of the central pressure are necessary
to get a potential barrier with a height similar to the one obtained when neutrinos are absent.
This means that for a given value of the central pressure the nucleation time will be much larger when neutrinos are present, and therefore,
the conversion of a HS to a QS will be inhibit in this case (see Figs.\ \ref{fig:figure2}
and \ref{fig:figure3}).

The nucleation time can be plotted as a function of the baryonic mass of the HS corresponding to the
given value of the central pressure, as implied by the solution of the Tolman-Oppenheimer-Volkov equations for the pure
HS sequences. The results of our calculation for the two cases (neutrino free and trapped) are reported in Fig.\ \ref{fig:figure2} for the same values of
$B$ and $\sigma$ of Fig.\ \ref{fig:figure1}. As it can be seen from the figure a metastable HS can have a mean-life time (related to the nucleation time)
many orders of magnitude larger than the age of the universe. Nevertheless, as the star accreates a small amount of mass ($\sim$ few percent of $M_\odot$),
the consequential increase of the central pressure leads to a huge reduction of the nucleation time and, as a result, to a dramatic reduction of the HS
 mean-life time. From the discussion of the previous figure, it is clear that if neutrinos are trapped the star will need to have a much larger baryonic mass
(or central pressure) than the star without neutrinos to have a comparable mean-life time. In the particular case of a star with a mean-life
time $\tau = 1$ year the baryonic mass of the star with neutrinos is $M_B \sim 2.12 M_\odot$, whereas if the neutrinos have diffused out of
the star  $M_B \sim 1.61 M_\odot$.

Finally, in Fig.\ \ref{fig:figure3} we show the mass-radius (MR) relation for the initial HS and final
QS sequences for the neutrino free (left panel) and neutrino trapped (right panel) cases. In both panels the configuration marked with an asterisk on the
hadron MR curves represents the HS for which the central pressure is $P_0$. All the configurations below this point
correspond to absolutely stable HS ($\tau = \infty$).
The full circles (triangle) on the HS and QS sequences of the neutrino free (trapped) case represent the HS for which $\tau = 1$ year
and the final QS which is formed from its conversion (in the neutrino trapped case deconfinement leads to the formation of a black hole for this particular
set of parameters). The square in the neutrino trapped case corresponds to the HS configuration which will evolve (assuming baryon number conservation) to the HS with $\tau = 1$
year in the neutrino free case as the neutrinos diffuse out of the star. It is clear from the figure that this and almost all the HS configurations
in the neutrino trapped case will be stable with respect to the conversion.

                                                                                                                                                                                                                 
\section{Conclusions}
\label{sec:sec4}
                                                                                                                                                                                                                 
In the present work we have studied the role of neutrino trapping in the HS to QS conversion mechanism
proposed in Ref. \cite{be03}. We have found that the quantum nucleation of a quark matter drop inside hadron matter,
and therefore the conversion of a HS into a QS, is strongly inhibit by the presence of neutrinos.

\end{document}